\documentclass{article}

\usepackage{amssymb}
\usepackage{amsfonts}
\usepackage{amsmath}

\usepackage{epsfig}
\usepackage{graphicx}
\usepackage{array}

\usepackage{graphics}

\def\piplus        {$\pi^{+}$}
\def\pimin         {$\pi^{-}$}

\def\kzeros        {$K^{0}_{S}$}

\def\lambdazero    {$\Lambda$}
\def\antilambda    {$\bar{\Lambda}$}
\def\ap    {$\bar{p}$}

\def\to    {$\rightarrow$}

\def\munit{\ifmmode{\,\mathrm{MeV/{\mit c}^{\,2}}}
          \else{MeV/$c^{\,2}$}\fi}
\def\mup{\ifmmode{\mathrm{\,MeV/{\mit c}}}
          \else{MeV/{\it c}}\fi}
\def\mupp{\ifmmode{\mathrm{\,(MeV/{\mit c})^2}}
          \else{(MeV/{\it c})$^2$}\fi}

\def\gunit{\ifmmode{\,\mathrm{GeV/{\mit c}^{\,2}}}
          \else{GeV/$c^{\,2}$}\fi}
\def\pup{\ifmmode{\mathrm{\,GeV/{\mit c}}}
          \else{GeV/{\it c}}\fi}
\def\pupp{\ifmmode{\mathrm{\,(GeV/{\mit c})^2}}
          \else{(GeV/{\it c})$^2$}\fi}
\def\pum{\ifmmode{\mathrm{\,(GeV/{\mit c})^{-1}}}
          \else{(GeV/{\it c})$^{-1}$}\fi}
\def\pumm{\ifmmode{\mathrm{\,(GeV/{\mit c})^{-2}}}
          \else{(GeV/{\it c})$^{-2}$}\fi}

\def\ptt  {$p^2_t$}
\def\xf   {$x_{F}$}

\def\to {$\rightarrow$}

\def\cpc#1#2#3  {{\rm Computer Phys. Comm.}  {\bf#1}, (#3) #2}
\def\npb#1#2#3  {{\rm Nucl. Phys. B}         {\bf#1}, (#3) #2}
\def\plb#1#2#3  {{\rm Phys. Lett. B}         {\bf#1}, (#3) #2}
\def\prd#1#2#3  {{\rm Phys. Rev. D}          {\bf#1}, (#3) #2}
\def\prl#1#2#3  {{\rm Phys. Rev. Lett.}      {\bf#1}, (#3) #2}
\def\sjnp#1#2#3 {{\rm Sov. J. Nucl. Phys.}   {\bf#1}, (#3) #2}
\def\spjl#1#2#3 {{\rm Sov. JETP Lett.}       {\bf#1}, #2 (#3)}
\def\zpc#1#2#3  {{\rm Z. Phys. C}            {\bf#1}, (#3) #2}

\def\rkl {$d\sigma(K^{0}_{S})/d\sigma(\Lambda)$}
\def\rll {$d\sigma(\bar{\Lambda})/d\sigma(\Lambda)$}
\def\rklpa {$d\sigma_{pA}$(K$^{0}_{S})/d\sigma_{pA}(\Lambda)$}
\def\rllpa {$d\sigma_{pA}(\bar{\Lambda})/d\sigma_{pA}(\Lambda)$}

\begin{document}

\pagestyle{empty}

\begin{flushright}
DESY-02-213 \\*[2mm]
Hep-ex/0212040 \\*[8mm]
\today
\end{flushright}
\vspace{0.8cm}
\Large
\centerline{
\parbox{12cm}{
\begin{center}
Inclusive $V^0$ Production Cross Sections
from 920\,GeV Fixed Target Proton-Nucleus Collisions
\end{center}
}}

\large
\vspace{1.8cm}
\centerline{The HERA-B Collaboration}
\vspace{0.4cm}
\normalsize
%
\begin{center}
I.~Abt$^{28}$,
A.~Abyzov$^{26}$,
M.~Adams$^{11}$,
H.~Albrecht$^{13}$,
V.~Amaral$^{8}$,
A.~Amorim$^{8}$,
S.~J.~Aplin$^{13}$,
A.~Arefiev$^{25}$,
I.~Ari\~no$^{2}$,
M.~Atiya$^{36}$
% ...
\\

\vspace*{0.7cm}
% ...
$^2$ {\sl Department ECM, Faculty of Physics, University of Barcelona, 
       E-08028 Barcelona, Spain} \\ 
% ...
$^8$ {\sl LIP Coimbra and Lisboa, P-3004-516 Coimbra, Portugal}  \\
% ...
$^{11}$ {\sl Institut f\"ur Physik, Universit\"at Dortmund, D-44227 Dortmund, 
       Germany~$^{\rm c}$}  \\
$^{12}$ {\sl Northwestern University, Evanston, Il 60208, USA$^{\rm a}$} \\ 
$^{13}$ {\sl DESY, D-22603 Hamburg, Germany}  \\ 
% ...
$^{25}$ {\sl Institute of Theoretical and Experimental Physics, 117259 Moscow,
       Russia~$^{g}$ } \\
$^{26}$ {\sl Joint Institute for Nuclear Research Dubna, 141980 Dubna, Moscow region,
       Russia} \\ 
% ...
$^{28}$ {\sl Max-Planck-Institut f\"ur Physik, Werner-Heisenberg-Institut, 
       D-80805 M\"unchen, Germany~$^{\rm c}$}  \\
% ...
$^{36}$ {\sl Brookhaven National Laboratory, Upton, NY 11973, USA} \\ 
% ...
$^a${\sl supported by the U.S. Department of Energy (DOE)} \\ 
% ...
$^c${\sl supported by the Bundesministerium f\"ur Bildung
       und Forschung, FRG, under contract numbers 05-7BU35I, 05-7DO55P, 
       05 HB1HRA, 05 HB1KHA, 05 HB1PEA, 05 HB1PSA, 05 HB1VHA, \\ 05 HB9HRA,
        05 7HD15I, 05 7HH25I,  05 7MP25I, 05 7SI75I}\\
% ...
$^g${\sl supported by the National Academy of Science and 
       the Ministry of Education and Science of Ukraine}\\
% ...

\end{center}
\vspace*{0.7cm}
\hrule
\vspace*{0.5cm}

\noindent{\bf Abstract} \\
Inclusive differential cross sections $d\sigma_{pA}/dx_F$ and $d\sigma_{pA}/dp_t^2$ 
for  the production of \kzeros, \lambdazero, and \antilambda\ particles are
measured at HERA in proton-induced reactions on C, Al, Ti, and W targets. The incident 
beam energy is 920~GeV, corresponding to $\sqrt {s} = 41.6$~GeV in the proton-nucleon 
system. The ratios of differential cross sections \rklpa\ and \rllpa\ are measured 
to be $6.2\pm 0.5$ and $0.66\pm 0.07$, respectively, for \xf\,$\approx\!-0.06$.
No significant dependence upon the target material is observed.
Within errors, the slopes of the transverse momentum distributions 
$d\sigma_{pA}/dp_t^2$ also show no significant dependence upon the target material. 
The dependence of the extrapolated total cross sections $\sigma_{pA}$ on the 
atomic mass $A$ of the target material is discussed, and the deduced cross sections 
per nucleon $\sigma_{pN}$ are compared with results obtained at other energies.
\vspace{1cm}

\noindent $Key\ words$: \ \ \  Hyperons, cross section, A-dependence

\noindent PACS: \ \ \  {{13.85.-t} \ \ \  {13.85.Ni} \ \ \  {13.85.Lg}
               } 
\vspace*{0.5cm}
\hrule

\clearpage

%
\section{Introduction}
\label{intro}

We present measurements of inclusive production cross sections 
for \kzeros, \lambdazero, and \antilambda\ particles,
collectively referred to as $V^0$ particles,
in collisions of 920~GeV protons with several nuclear targets of different
atomic mass~$A$. The production of $V^0$ particles has been studied by
numerous experiments with different beams ($\pi$, $K$, $p$, $n$, $\Sigma^-$)
on different targets, covering a  momentum range of 4--400 \pup\ (see 
\cite{biswas80,mikocki86,ljung77,bogert77,edwards78,oh72,blobel74,jaeger75,boggild73,erwin75,chapman73,jaeger751,sheng75,heller77,skubic78,dao73,asai85,kichimi78,aleev86,v0wa89} 
and refs. therein). 
A large fraction of these experiments utilized bubble chambers. 
At center-of-mass energies above $\sqrt {s} \approx 30$~GeV, inclusive 
$V^0$ production cross sections have been measured by experiments at 
the CERN proton-proton Intersecting Storage Rings 
(ISR)~\cite{busser76,erhan79,drij81,drij82}. 
However, only a lower limit for the \antilambda\ cross section 
has been reported~\cite{erhan79}.
Also, the $V^0$ total production cross sections reported in \cite{drij82} 
are substantially above expectations based on extrapolation of results 
obtained at lower energies.

Recently, progress in heavy ion physics has renewed interest in studies 
of strange-particle production.  One of the main goals of heavy-ion experiments
is the observation of the quark-gluon plasma~\cite{RHIC}, and one of the 
signatures for this new state is enhanced 
production of strange particles~\cite{RHIC99}.  
Observables of interest are ratios of antibaryons to baryons at mid-rapidity, 
which are important for net baryon density evaluations and which have been used 
recently to extract values of chemical potentials and 
temperatures~\cite{raf01,brm01} for Au-Au collisions at RHIC. 
Also of interest are the squared  transverse momentum (\ptt ) distributions, 
$d\sigma/d$\ptt, which contain information about the temperature of the system; 
specifically, if it reached thermal equilibrium (see, e.g.,~\cite{brav99}).
These investigations attach importance to a comprehensive measurement of 
strange-particle production properties in ``ordinary'' nucleon-nucleon ($NN$) 
and nucleon-nucleus ($NA$) collisions.  The latter are expected to establish 
a valuable baseline for comparisons among $AA$ results~\cite{raf01}. 

\section{The HERA-B Experiment}
\label{sec:exp}
HERA-B is a fixed target experiment at the 920~GeV proton 
storage ring of HERA at DESY~\cite{HERAB95}. 
It was originally designed to study CP violation in 
$B^0 \rightarrow J/\psi K^0_S$ decays; thus,
the spectrometer was optimized to detect and precisely
locate decays of $J/\psi$ and longer-lived
$K^0_S$ mesons. The first experience with the apparatus was gained 
during an engineering run in the year 2000.
While the main goal was to commission the sophisticated $J/\psi$-based
trigger system, there were also run periods dedicated to recording
minimum bias events, i.e., events selected with a random trigger 
that uniformly sampled all HERA bunches. These data allow for precise 
measurements of strange-particle production cross sections.

\begin{figure*} 
\addtolength{\abovecaptionskip}{10pt}
\centering
\includegraphics[width=17cm]{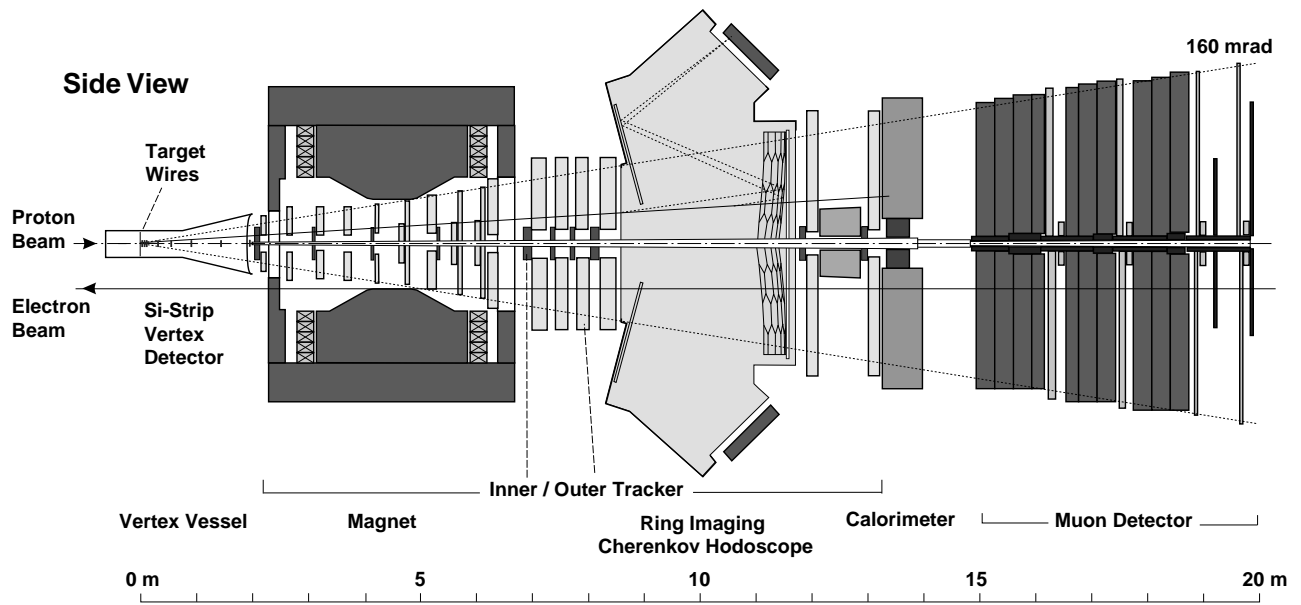}
\caption{Plan view of the HERA-B detector.}
\label{fig:herab}
\end{figure*}

A plan view of the HERA-B spectrometer is shown in  Fig.~\ref{fig:herab}. 
The spectrometer dipole magnet provides a field integral of 2.13~T-m, with the 
main component perpendicular to the $x$-$z$ plane. 
The apparatus (including particle identification
counters) has a forward acceptance of 10--220~mrad in the bending 
plane and 10--160~mrad in the non-bending plane.
The experiment uses a multi-wire fixed target which operates in the halo of 
the proton beam during HERA $e$-$p$ collider operation. 
Up to eight different targets can be operated simultaneously, with their 
positions being adjusted dynamically in order to maintain a constant interaction 
rate between 1 and 40~MHz. 
For the measurements presented here, wires made of carbon (C), aluminum (Al), 
titanium (Ti), and tungsten (W) were used. Their transverse width was 50~$\mu $m, 
and their thickness along the beam direction was 1000~$\mu $m for carbon and 
500~$\mu $m otherwise.

The tracking system consists of a vertex detector system (VDS)
and a main tracker system. The latter is separated into an inner 
tracker (ITR) close to the proton beam-pipe and an outer tracker (OTR)
farther out.
The VDS~\cite{vds} features 64 double-sided silicon microstrip detectors
arranged in eight stations along, and four quadrants around, 
the proton beam. 
The silicon strips have a readout pitch of approximately 50~$\mu $m. 
The strips are rotated on the wafer such that each pair of wafers 
provides four stereo views: $\pm 2.5^\circ$ and $90^\circ \pm 2.5^\circ$. 
The ITR uses GEM microstrip gas chambers~\cite{itr}, and
the OTR uses honeycomb drift chambers~\cite{otr}.
There are a total of 13 ITR\,+\,OTR tracking stations,
with each one referred to as a ``superlayer.'' For the analysis
presented here, only data from the six superlayers located 
downstream of the magnet are used. Also, this analysis does 
not use particle-identification information from the ring-imaging 
Cherenkov counter~\cite{rich}, the electromagnetic 
calorimeter~\cite{ecal}, or the muon detector~\cite{muon}.

\section{Data Analysis}
\label{sec:evs}

The results presented here are based on a sample of
$\sim$\,2.4 million randomly triggered events recorded 
in April 2000. 
The HERA proton beam had 180 filled bunches 
and a bunch crossing rate of 10.4~MHz.
The interaction rate was adjusted to values between 2 and 6 MHz,   
leading to most bunch crossings ($\geq 90\%$) having either 
0 or 1 interaction.
For this analysis, only tracks with a minimum of five hits 
in the VDS and ten hits in the main tracker were used.
The same event selection procedures were applied as were
used to determine the integrated luminosity. No explicit selection 
on track multiplicity was applied.

\begin{figure}[htb] 
\centering
\includegraphics[width=11cm]{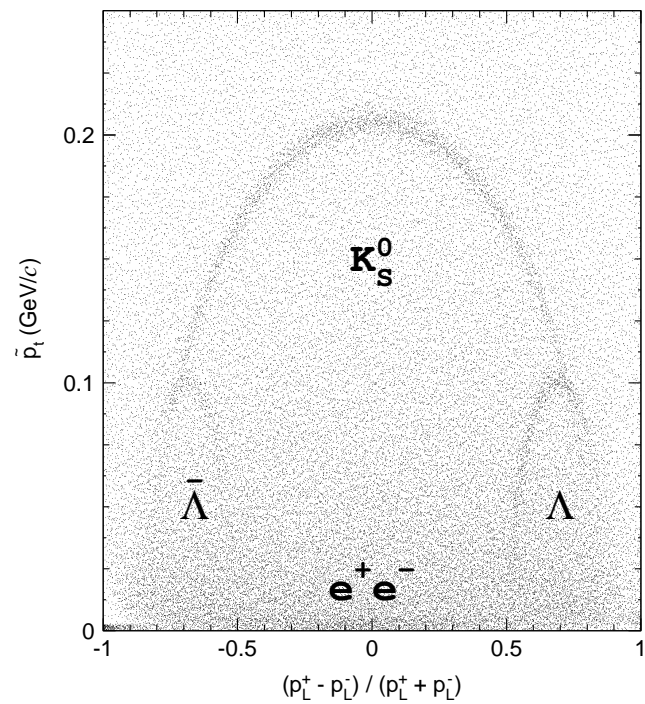}
\caption{The Armenteros-Podolanski plot for the $V^0$
         candidates: the transverse momentum $\tilde{p}_t$ of 
	the oppositely charged decay products  vs. their
	 asymmetry in longitudinal momenta $p^{\pm}_L$.
	All momenta are relative to the $V^0$ line-of-flight.  
	Background from $\gamma\!\rightarrow\!e^+ e^-$ conversions
	populates the region below $\tilde{p}_t = 0.015~\pup $.} 
\label{fig:armentero}
\end{figure}

In each event with at least two tracks, a full combinatorial search 
for $V^0$ candidates was performed. $V^0$ candidates were selected 
from all pairs of oppositely charged tracks that formed a vertex 
downstream of the primary vertex.
The minimum distance between the two tracks was required to be less 
than 0.7~mm. No particle identification criteria were applied to 
the tracks. The primary vertex was determined from all 
tracks reconstructed in the VDS excluding the $V^0$ tracks.
Using tracks reconstructed in both the VDS and the main tracker, 
we measured the spatial resolution of the reconstructed primary 
vertex to be 0.7~mm along the $z$-direction.
The position of the primary vertex was required to coincide with the 
center of a target wire to within three standard deviations (2.1~mm). 
If a primary vertex could not be reconstructed, the $z$ coordinate of the 
target was used to calculate the $z$-component of the particle's flight length.
An Armenteros-Podolanski plot~\cite{ARPP54} for the selected 
$V^0$ candidates is shown in Fig.~\ref{fig:armentero}. 
Clusters of events  shaped according to the kinematics of 
\kzeros\,\to\,\piplus\,\pimin , \lambdazero\,\to\,$p$\,\pimin\ and  
\antilambda\,\to\,\ap\,\piplus\ decays are clearly visible.
The $V^0$ candidates from those regions of Fig.~\ref{fig:armentero}
in which two of the three $V^0$ species overlap --\,and thus are kinematically 
indistinguishable\,-- were rejected. This reduced the \kzeros\ yield by 3.5\% 
and the \lambdazero\  (\antilambda ) yield by 10\% (20\%).
An additional requirement $\tilde{p}_t\cdot c\tau > 0.05~(\pup )\cdot {\rm cm}$
was applied to reduce background from $\gamma\!\rightarrow\!e^+e^-$ 
conversions, where $\tilde{p}_t$
is the transverse momentum relative to the $V^0$ 
line-of-flight and $\tau$ is the $V^0$ proper lifetime.
This requirement also reduced 
combinatorial background from the target, which populates 
the lower region of Fig.~\ref{fig:armentero}.

\begin{figure}[htb] 
\centering
\includegraphics[width=11cm]{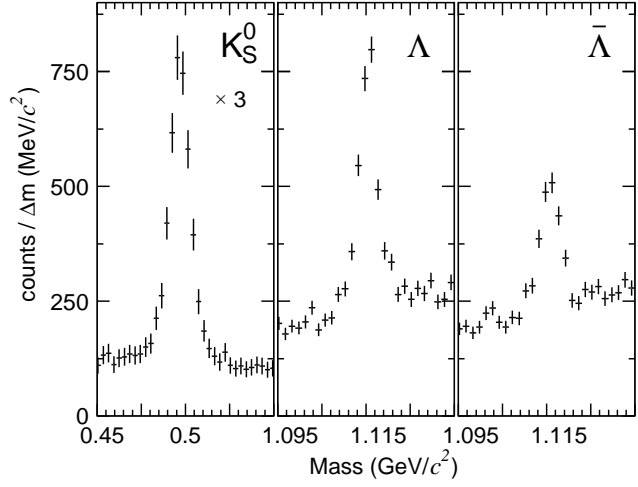}
\caption{ The invariant mass distributions for \kzeros, \lambdazero, and 
\antilambda\ particles summed over all runs with different targets. The 
size of the invariant mass bins ($\Delta m$) is 3.0~MeV/{\it c$^2$} for 
the \kzeros\ distribution and 1.5~MeV/{\it c$^2$} otherwise.
}
\label{fig:v0mass}
\end{figure}

The invariant mass distributions for the selected candidates 
are shown in Fig.\,\ref{fig:v0mass}. 
Clear signals corresponding
to  \kzeros , \lambdazero, and \antilambda\  particles are visible.
In order to estimate the event yields,
fits with a Gaussian function for signal and a third-order Legendre
polynomial for background were performed.
The width obtained for the \kzeros\ mass peak was about
5.6~MeV/{\it c$^2$}, which is consistent with the momentum 
resolution of $\sigma(p) / p^2 \approx 10^{-4}$ \pum.
The resultant event yields are summarized in 
Table~\ref{tab:counts} for each target material.

\section{Acceptance Determination}
\label{sec:accept}

A Monte Carlo (MC) simulation is used to determine the 
reconstruction efficiency for the selected particles
and decay channels. 
The production of a $V^0$ particle in an inelastic event 
is simulated using a kinematic distribution of the form:
\begin{equation} 
\frac{d^2\sigma}{dp^2_t dx_F} = C\cdot(1 - |x_F|)^n \cdot exp (-B\cdot p_t^2)\,,
\label{eq:xfpt}
\end{equation}
where $x_F$ is the Feynman-$x$ variable and $p_t$ the transverse momentum 
in the laboratory system. This phenomenological ansatz is motivated by 
quark counting rules and phase space arguments~\cite{brodsky78} and has 
been shown to describe a substantial amount of data well 
(see \cite{v0wa89} and refs.~therein).
The value of the parameter $B$ of 2.1\,(GeV/$c$)$^{-2}$ is taken 
from Ref.~\cite{v0wa89}, and a flat $x_F$ distribution is assumed 
($n\!=\!0$). The \xf\ bin size of 0.015 is chosen to be 
compatible with the momentum resolution.  
After the generation of the $V^0$ particle, the remaining momentum is assigned 
to a virtual \piplus, which is input as a beam particle 
to the FRITIOF~7.02 package~\cite{fritiof} in order to 
further simulate interactions within the nucleus. 

The generated particles are propagated through the geometry and
material description of the detector using the GEANT 3.21 
package~\cite{geant}. 
The detector response is simulated including realistic descriptions 
of chamber efficiencies and dead channels. 
The MC events are subjected to the same reconstruction
chain as that used for the data.
For tracks within the geometrical acceptance that originate from $V^0$
decays, we determine an efficiency of $\sim$\,90\% 
to reconstruct the track and assign a momentum.  
This value is confirmed by the analysis of real data.

After all cuts, the efficiency to identify a $V^0$ particle is 
approximately 10\% for \kzeros \ and 5\% for \lambdazero\ and \antilambda.
These values are dominated by the geometrical acceptance. All efficiencies 
are computed in bins of rapidity $y$ and \ptt\ in order to be independent 
of the details of the MC model of $V^0$ production.

\section{ Production Cross Sections}
\label{sec:xsect}

The cross section $\Delta \sigma_{V^0}$ for the production of a $V^0$ 
particle within the spectrometer acceptance can be expressed as: 

\begin{equation}
      \Delta \sigma_{V^0}  = {\frac { 1 }  { B_{V^0}\ {\mathcal L} }} \iint_{accp}
       {\frac { N_{V^0}(y,p^2_t) }  {\varepsilon(y,p^2_t)
       } } dy\cdot dp^2_t\,,
\label{eq:crsec}
\end{equation}
where $N_{V^0}(y,p^2_t)$ is the number of $V^0$ candidates
observed in bins of rapidity $y$ and \ptt.
The detection efficiency $\varepsilon(y,p^2_t)$ is calculated
from the MC simulations described above.
The branching fraction $B_{V^0}$ for the 
detected decay $V^0\!\rightarrow\!f$ is taken from Ref.~\cite{PDG}.
Finally, the integrated luminosity ${\mathcal L}$ is computed 
from the interaction rate at the target~\cite{lumi}. 
The resultant luminosities for each target material are listed 
in Table~\ref{tab:counts}.

The efficiency-corrected values of the inclusive differential 
cross section $d\sigma_{pA}/dx_F$ for $V^0$ production 
within the spectrometer acceptance are 
listed in Table~\ref{tab:csection} for the various target materials. 
The corresponding \xf\ interval is $[-0.12, 0\,]$, 
except in the case of $K^0_S$ production on the W target, where 
it is  $[-0.09, 0\,]$.

As the \xf\ acceptance of the spectrometer is restricted, 
for comparison with results from other experiments it is necessary 
to~extrapolate~~the~~differential~cross~sections $d\sigma_{pA}/dx_F$ 
to the entire kinematic range \xf\,$\in [-1, +1\,]$. This was 
done using the parameterization ${d\sigma}/{dx_F} \propto (1 - |x_F|)^n $,
where $n$ is a constant.
The values used for $n$ are taken from previous measurements of inclusive 
strange-particle production~\cite{v0wa89} and are listed in Table~\ref{tab:csection}.
The difference between values for different $V^0$ species is compatible with 
theoretical predictions based upon quark counting rules~\cite{brodsky78}.
The resulting $V^0$ total production cross sections ($\sigma_{pA}$) 
are also listed in Table~\ref{tab:csection}. The fraction of the cross 
sections within the detector acceptance is 30\% for \kzeros \ ($n=6.0$), 
17\% for \lambdazero\ ($n=2.2$), and 34\% for \antilambda \ ($n=8.0$).

\begin{figure} 
\centering
\includegraphics[width=9cm]{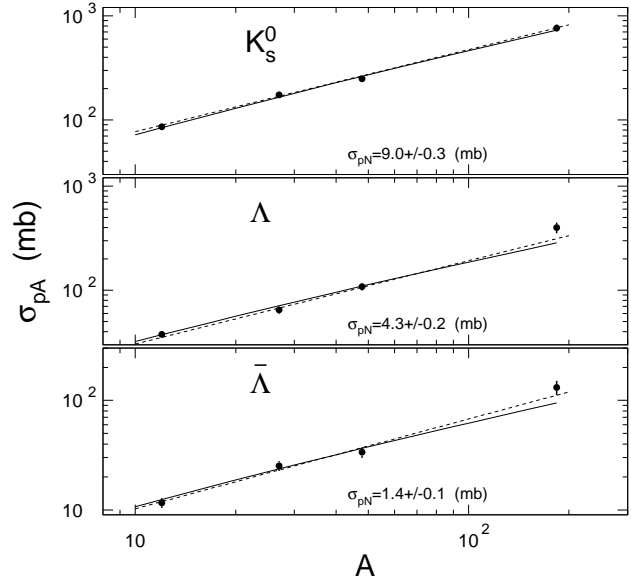}
\caption{The $V^0$ total production cross sections $\sigma_{pA}$ as a 
  	 function of the atomic mass $A$ of the target material. 
         The solid lines show fits within the Glauber model, which yield 
	 the indicated production cross sections per nucleon 
         ($\sigma_{pN}$). 
         The dashed lines are fits to the form
         $\sigma_{pA} \propto A^{\alpha}$; the resultant
         $\alpha $ values are listed in Table~\ref{tab:nsection}.
         }
\label{fig:cs_glauber}
\end{figure}

The dependence of the measured cross sections on the atomic mass of the target 
material is shown in Fig.\,\ref{fig:cs_glauber} along with fits of two different 
kinds. The dashed lines are fits to the form $\sigma_{pA} \propto A^{\alpha}$, 
while the solid lines are fits within the framework of the 
Glauber model~\cite{glauber59}. These latter fits allow us 
to extract the production cross section per nucleon, $\sigma_{pN}$. All fit 
results are listed in Table~\ref{tab:nsection}. The Glauber model calculations 
use a nucleon density given by the Saxon-Woods potential \cite{denisov73},
and the total cross sections for $KN$ and $\Lambda N$ collisions and the slope 
of the diffraction scattering cone are taken from Refs.~\cite{denisov73,PDG}.

\subsection{ The \ptt\ Differential Cross Sections}
\label{sec:diffxsect}
\begin{figure}[htb] 
\centering
\includegraphics[width=10cm]{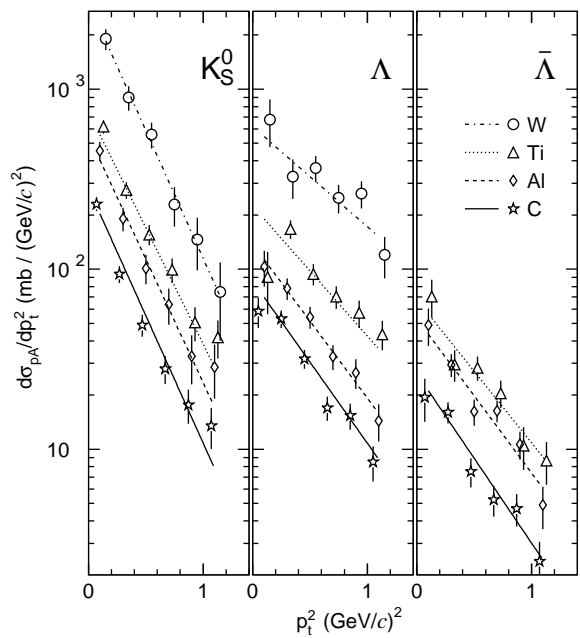}
\caption{The differential cross section $d\sigma_{pA}/dp_t^2$ for \kzeros, 
\lambdazero, and \antilambda\ production on indicated target materials
as obtained by extrapolation to the full $x_F$ range $[-1, +1\,]$.
}
\label{fig:dndpt2}
\end{figure}

The extrapolation of the the \ptt\ distributions for \kzeros , \lambdazero, and 
\antilambda\ particles over the entire $x_F$ range $[-1, +1\,]$ yields the 
differential cross sections $d\sigma/dp^2_T$ listed in 
Table\,\ref{tab:diffptall}. 
These data are presented graphically in Fig.\,\ref{fig:dndpt2} together 
with fits to the form
\begin{equation}
\frac{d\sigma}{dp^2_t} = \sigma\cdot B\cdot exp (-B\cdot p_t^2)\,, 
\label{eq:param}
\end{equation}
where the parameter $B$ is independent of \xf\ and \ptt . 
Table~\ref{tab:bparam} summarizes the values of $B$ obtained for 
the different target materials and $V^0$ particles. No significant 
dependence upon the target material is seen. 

\subsection{ Systematic Uncertainties}

To estimate the uncertainty in the efficiency determination, we varied the 
$V^0$ selection criteria applied to data and simulated events.  
Different track quality requirements used in the vertex reconstruction were
studied in order to understand differences observed between data and MC
results; the variation of cuts causes a change of 7\% in the cross section 
values.
Cuts on the primary and $V^0$ vertices give a 2\% change.
The variation of the distance required between tracks to form a $V^0$ decay vertex 
contributes another 2\%.
The systematic error due to the limited statistics of  MC events
is 1\% for \kzeros\ and 3\% for \lambdazero \ and \antilambda \ particles.
The use of different fitting functions for the shape of the background in
the invariant mass spectra, and the variations of bin sizes for invariant 
mass, $x_F$, and \ptt, result in a change in the cross sections of 
approximately~5\%.

The efficiency of the ITR improved during data taking (as this system was
further commissioned), and this caused a systematic shift in the 
track reconstruction efficiencies. These  shifts
are +14\%, +11\%, and +20\% for \kzeros, \lambdazero, and 
\antilambda\ particles, respectively.

We used three independent methods to determine the integrated luminosity.
The results differ by 4--7\%, which we include as an additional
systematic uncertainty. For more details, see Ref.~\cite{lumi}. 

To extrapolate $d\sigma/dp^2_t$ to the entire kinematic range of \xf,
we used Eq.~\ref{eq:xfpt} with experimentally determined values~\cite{v0wa89} 
of the parameter $n$. Varying this parameter by the 
estimated experimental uncertainty ($\pm 0.5$) leads to
a variation of 5\% in the \kzeros \ cross sections, 12\% in the
\lambdazero\ cross sections, and 2\% in the \antilambda \ cross sections.

The cross sections were measured on different nuclei, and extracting
the cross sections per nucleon ($\sigma_{pN}$) from the measured values 
was done via the Glauber model. Variations of the parameters of this model
give a 12\% systematic uncertainty in $\sigma_{pN}$.

Adding all systematic errors in quadrature, we obtain the total 
systematic errors for the cross sections per nucleus
$d\sigma_{pA}/dx_F$, $d\sigma_{pA}/dp_t^2$, and $\sigma_{pA}$,
and for the cross section per nucleon $\sigma_{pN}$. All systematic
errors are given in Tables~\ref{tab:csection}, \ref{tab:nsection},
and \ref{tab:diffptall} (listed after statistical errors); the errors
are asymmetric due to changes in running conditions for the ITR. 

\section{ Discussion of Results}
\label{sec:disc}
\begin{figure}[h,t] 
\centering
\includegraphics[width=10cm]{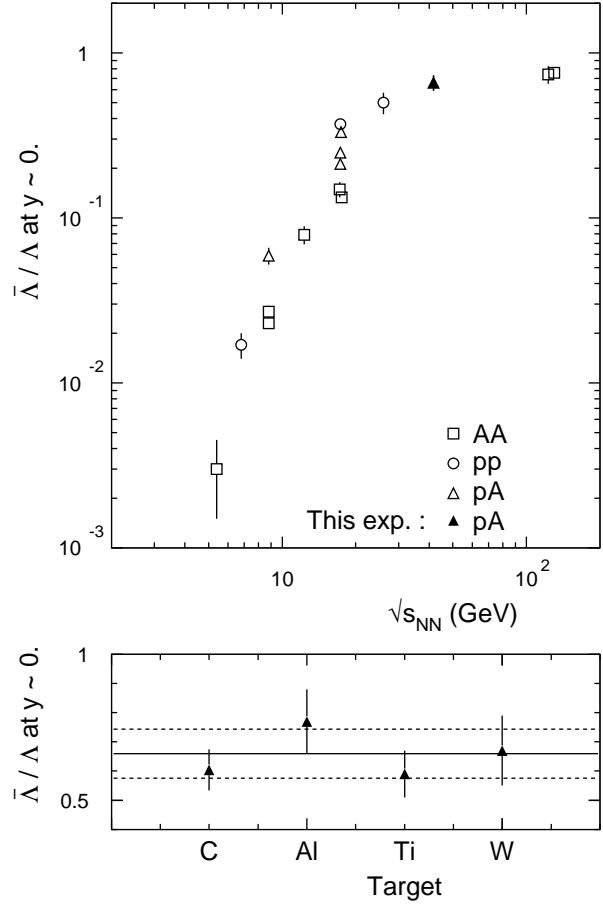}
\caption{
Top: The ratio of $\bar{\Lambda}$ and $\Lambda$ particle yields at 
mid-rapidity 
for $pp$ (circles), $pA$ (triangles), and $AA$ (squares) collisions as a
function of the nucleon-nucleon center-of-mass energy $\sqrt{s_{NN}}$.
The black triangle denotes the average of the HERA-B results; references 
for the open symbols are given in the text.
Bottom: The ratio \rllpa\ determined at $x_F\approx -0.06$ for the 
indicated 
targets; the average of the four values and the standard deviation are 
indicated by the solid and dashed lines, respectively. 
} 
\label{fig:llbar}
\end{figure}

More precise than the production cross sections themselves are cross section ratios,
since, for these, acceptance corrections and systematic errors to a large
extent cancel. From the measurements presented above we calculate the ratios 
\rkl\ = $6.2\pm 0.5$ and \rll\ = $0.66\pm 0.07$, for $x^{}_F\approx -0.06$ 
(i.e., mid-rapidity). 
The \rll\  ratio is plotted in Fig.\,\ref{fig:llbar} for 
the different target materials; no significant dependence upon the  
target material (i.e., atomic mass $A$) is seen, and this holds also for the \rkl\ ratio.

Ratios between antiparticle yields and particle yields measured at 
mid-rapidity have attracted special interest in studies of the dependence 
of $AA$ collision dynamics upon collision energy~\cite{raf01,brm01}.
Fig.\,\ref{fig:llbar} shows the ratio of yields for
$\bar{\Lambda}$ and $\Lambda$ particles at mid-rapidity 
in $AA$ collisions as a function of center-of-mass energy $\sqrt{s_{NN}}$.
The data plotted are from the BNL AGS ($\sqrt{s_{NN}}\!\approx\!5.4$~GeV), 
the CERN SPS ($\sqrt{s_{NN}}\!\approx$\,10--30~GeV), and 
RHIC ($\sqrt{s_{NN}} = 130$~GeV) 
\cite{kadija02,fanebust02,caliandro99,cole95,mischke02,adcox02,adler02}.
Comparing these measurements with corresponding results from 
$pp$ \cite{blobel74,asai85,kadija02} 
and
$pA$ \cite{kadija02,fanebust02,caliandro99}
collisions shows these data to exhibit a similar energy dependence, 
and the HERA-B result at $\sqrt{s} = 41.6$~GeV
is consistent with this trend. Closer inspections suggests that
the $pp$, $pA$, and $AA$ data follow individual curves that
are slightly shifted in energy. This would be consistent with results
from NA49~\cite{kadija02}, which found that in $pp$ collisions 
at $\sqrt{s_{NN}}\!=\!17.3$~GeV the ratio \rll\ is about 60\% larger 
than that in $pA$ collisions, and in $AA$ collisions at the
same $\sqrt{s_{NN}}$ the ratio is about 60\% smaller.
However, more data are needed to confirm this behavior. 

The measured \ptt\ spectra presented above 
are well-parametrized by Eq.~\ref{eq:param}. 
The results (Table~\ref{tab:bparam})  
show no significant dependence upon the target material, 
and they are compatible with results from  
other experiments,
 e.g. from $pp$ collisions at $\sqrt{s} = 27.6$~GeV, where 
at \ptt \ $<$ 1.5~(GeV/$c$)$^2$ B parameters
for \kzeros , \lambdazero\ and \antilambda\ particles of 
$3.59\pm 0.18$~(GeV/$c$)$^{-2}$, 
$2.73\pm 0.21$~(GeV/$c$)$^{-2}$ and 
$2.66\pm 0.74$~(GeV/$c$)$^{-2}$, respectively, have been extracted 
\cite{kichimi78};
for more examples see the compilation in \cite{v0wa89}. 
Due to the limited statistics available, the present study was 
done in the \ptt \ region below 1.1~(GeV/$c$)$^2$; thus we 
cannot confirm the flattening of the \ptt\ spectrum above 
1.2~\pupp\ (see e.g.~\cite{asai85}).

The narrow acceptance in \xf\ for this data set also precludes  
us from measuring ``leading particle'' effects, which have been 
observed by other fixed target
experiments at $x_F\!>\!0.4$ (see e.g. \cite{v0wa89,basile81}). 
However, we have measured for the first time $V^0$ 
differential cross sections for different target 
materials at negative \xf\ ($-0.12\!<\!x_F\!<\!0$).
We extrapolate these results to the full
\xf\ range to obtain total cross sections
for different values of atomic mass~$A$. 
We fit these values to the conventional
expression $\sigma_{pA} \propto A^{\alpha}$ and also to 
the Glauber model.
The values obtained for $\alpha$ from the first fit (Table~\ref{tab:nsection}) are
very similar to those observed in other hadroproduction processes, 
which can be characterized~\cite{GEI91} by
$\alpha (x_F) = 0.8 - 0.75\,x_F + 0.45\,x_F^3/|x_F|$\,. 

\begin{figure}[h] 
\centering
\includegraphics[width=10cm]{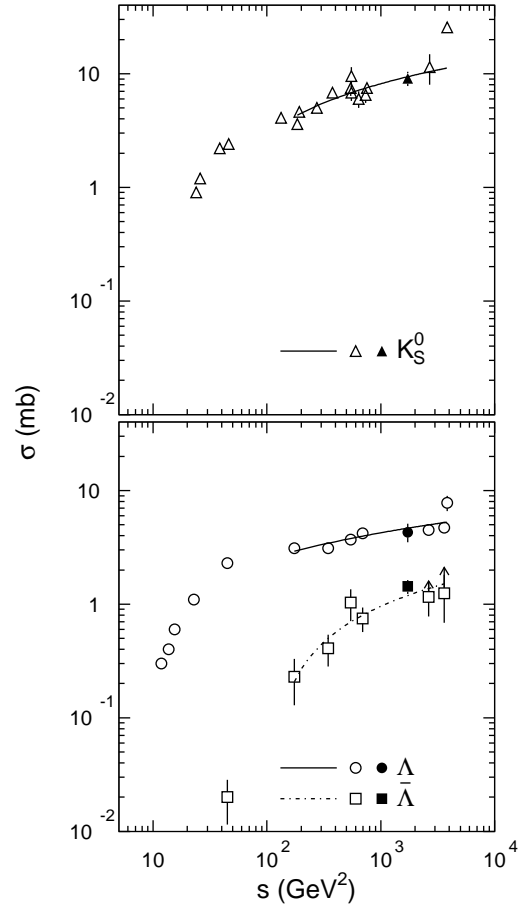}
\caption{The total cross sections per nucleon $\sigma_{pN}$ for the 
production of \kzeros, \lambdazero,  and 
\antilambda \ particles as a function of $s$, the square of the 
center-of-mass energy. Black symbols denote the results from  HERA-B, 
open symbols those from 
\cite{oh72,blobel74,jaeger75,boggild73,erwin75,chapman73,jaeger751,sheng75,heller77,skubic78,dao73,asai85,kichimi78,aleev86,v0wa89,busser76,erhan79,drij81,drij82} 
and refs. therein. The curves are calculated using the fit functions 
reported 
in Ref.~\cite{kichimi78}.
}  
\label{fig:inclusive}
\end{figure}

From the Glauber-model fits we extract $V^0$ cross sections 
per nucleon, $\sigma_{pN}$ (Table~\ref{tab:nsection}). 
Fig.\,\ref{fig:inclusive} presents a comparison of these results
with previous experimental results
\cite{oh72,blobel74,jaeger75,boggild73,erwin75,chapman73,jaeger751,sheng75,heller77,skubic78,dao73,asai85,kichimi78,aleev86,v0wa89,busser76,erhan79,drij81,drij82}.
The HERA-B results show
good agreement with the general systematic trend as quantified by the curves. 
These curves were calculated from a fit~\cite{kichimi78} for the average number
of $V^0$ particles per inelastic collision as a function of $\ln(s)$, in the interval 
$13.5\!<\!\sqrt{s}\!<\!28$~GeV. The data in Fig.~\ref{fig:inclusive} suggest that 
this parametrization is valid up to the HERA-B energy of $\sqrt{s} = 41.6$~GeV, 
and even up to the highest ISR energy of $\sqrt{s} = 63$~GeV.  
Our value for the \kzeros\ cross section is in good agreement with the value 
obtained at $\sqrt{s}\!=\!52.5$~GeV \cite{drij81}.
The large cross section reported in \cite{drij82} for $\sqrt{s}\!=\!63$~GeV  has been
recognized \cite{as02} to be biased by ghost tracks.
In addition, our value for the \lambdazero\ total cross section 
per nucleon is in good agreement with the values obtained at the 
ISR at $\sqrt{s}\!=\!53$~GeV and $\sqrt{s}\!=\!62$~GeV~\cite{erhan79}. 
As $\Lambda$ production is known to receive large contributions from 
fragmentation processes,
such a good agreement is noteworthy: it implies that also for small \xf\ 
the parameterization of the (double-differential) cross section given in 
Eq.~\ref{eq:xfpt} is valid.

\section*{Acknowledgments}

We express our gratitude to the DESY laboratory for the strong support in
setting up and running the HERA-B experiment. We are also indebted to the
DESY accelerator group for the continuous efforts to provide good and
stable beam conditions. 
The HERA-B experiment would not have been possible without the enormous
effort and commitment of our technical and administrative staff. It is a
pleasure to thank all of them. \\
We thank the theoreticians B.~Kopeliovich and H.J.~Pirner for many stimulating 
discussions and suggestions. \\
%

%
\clearpage
\begin{table}
\caption{
Integrated luminosities ${\mathcal L}_A$ in mb$^{-1}$ for the indicated 
targets of atomic mass~$A$~\cite{lumi}, and the corresponding numbers of 
events $N(evt)$ and $V^0$ particles $N(V^0)$ obtained. 
Uncertainties listed for ${\mathcal L}_A$ are systematic; 
those listed for $N(V^0)$ are statistical.        
}
\begin{center}
\begin{tabular}{|l|c|c|c|c|} \hline
             & C &Al & Ti & W   \\ \hline 
$A$           &  12 & 27 & 48 & 184 \\ 
${\mathcal L}_A\cdot$mb & $1093\pm 38 $ & $1030\pm 67\ $ & $308 \pm 10 $ & $47\pm\ 3\ $ \\
\hline 
$N(evt)$            &  496694    & 893885  & 467943  & 512897  \\ 
$N(K^0_S )$      & $2566\pm58$   & $4398\pm81$  & $2022\pm52$  & $909\pm40$  \\
$N(\Lambda )$  & $\ 512\pm31$  & $\ 831\pm43$ & $\ 412\pm27$  & $224\pm24$ \\
$N(\overline{\Lambda} )$  & $\ 241\pm24$ & $\ 503\pm41$ & $\ 222\pm23$ & $144\pm22$ \\ 
\hline
\end{tabular}
\end{center}
\label{tab:counts}
\end{table}
%
\begin{table}
\caption{ The inclusive differential cross sections $d\sigma_{pA}/dx_F$ in mb
measured in the \xf\ interval $[-0.12, 0\,]$ for the production of \kzeros, 
\lambdazero, and \antilambda\ particles on indicated targets. Also listed
are the total cross sections $\sigma_{pA}$ obtained by 
extrapolation to the full $x_F$ range (see text). 
Values are followed by statistical and systematic errors.
The values of the parameter $n$ used for the extrapolations
(see Eq.~\ref{eq:xfpt}) are also given.  
The \xf\ interval for the case of $K^0_S$ production on 
the W target is $[-0.09,0\,]$.
}
\begin{center}
\begin{tabular}{|l|c|c|c|c|} \hline
        &Trgt &  $n$  &  $d\sigma_{pA}/dx_F$ (mb) & $\sigma_{pA}$ (mb)  \\ 
\hline 
$K^0_S$ & C\  &6.0
        & $\ 215.\pm \ 12._{- \ 13.}^{+\ 33.}$
        & $\ 86.\pm \  5._{- \ \ 7.}^{+\ 14.}$ \\
        & Al &
        &  $\ 429.\pm \ 33._{-\  34.}^{+\ 68.}$
        & $ 174.\pm  13._{-\  16.}^{+\ 30.}$  \\
        & Ti &
        &  $\ 612.\pm \ 33._{- \ 37.}^{+\ 92.}$
        & $ 248.\pm  13._{- \ 20.}^{+\ 40.}$ \\
        & W\  &
        & $2044.\pm167._{- 156.}^{+322.}$
        & $761.\pm 63._{- \ 68.}^{+129.}$ \\ 
\hline
\lambdazero & C\  &2.2
            & $\ 53.\pm\ 4._{-\ 3.}^{+\ 7.}$
            & $\ 38.\pm\ 3._{- \ 5.}^{+\ 6.}$ \\
            & Al &
            & $\ 90.\pm\ 8._{- \ 8.}^{+13.}$
            & $\ 64.\pm \ 6._{- \ 9.}^{+12}$  \\
            & Ti &
            & $163.\pm 13._{- 13.}^{+21.}$
            & $115.\pm 20._{-  15.}^{+20}$ \\
            & W\ &
            & $558.\pm 62._{- 45.}^{+78.}$
            & $399.\pm 47._{-  56.}^{+76}$ \\ 
\hline
\antilambda & C\ &8.0
            & $\ 32. \pm \ 3._{- \  2.}^{+\ 7.}$
            & $\ 11.\pm \ 1._{- \ 0.7}^{+\ 2.3}$ \\
            & Al &
            & $\ 69. \pm \ 8._{- \ 6.}^{+15.}$
            & $\ 25.\pm \ 3._{- \ 2.7}^{+\ 5.3}$  \\
            & Ti &
            & $\ 96. \pm 11._{- \ 6.}^{+20.}$
            & $\ 34.\pm \ 4._{- \ 2.7}^{+\ 7.5}$ \\
            & W\ &
            & $375.\pm 58._{- 33.}^{+83.}$
            & $131.\pm 20._{-  10.\ }^{+29.}$ \\ 
\hline
\end{tabular}
\end{center}
\label{tab:csection}
\end{table}

%
%
\begin{table}[h]
\caption{ The $V^0$ production cross sections per nucleon ($\sigma_{pN}$) 
        in mb with statistical and systematic errors, and the 
        values of $\alpha$ resulting from fitting the cross sections 
        per nucleus to the form $\sigma_{pA} \propto A^{\alpha}$. For 
        the values of $\alpha$, the errors listed are statistical only. 
        } 
\begin{center}
\setlength{\extrarowheight}{4pt}
\begin{tabular}{|l|c|c|c|} \hline
         & \kzeros & \lambdazero & \antilambda \\ \hline
$\sigma_{pN}$ (mb)   &$9.0\ \pm0.3_{- 1.3}^{+1.8} $
                              &$4.3\ \pm0.2_{- 0.7}^{+0.9} $
                              &$1.4\ \pm0.1_{- 0.2}^{+0.3} $ \\ 
\hline 
$\alpha                $&$0.78\pm0.04$&$0.85\pm0.05$&$0.82\pm0.07$ \\ \hline
\end{tabular}
\end{center}
\label{tab:nsection}
\end{table}
%
\begin{table*}
\caption{ The inclusive differential cross sections ${d\sigma_{pA}}/{dp^2_t}$  in 
          mb/(\pup)$^{2}$ for the
          production of $V^0$ particles on indicated targets as obtained from
          extrapolation to the full $x_F$ range.
          Values are followed by  statistical and systematic errors.
          The \ptt\ bins ($\Delta$\ptt) are in (GeV/{\it c})$^2$; 
          the $p_t$ resolution is 1.7~MeV/c.
          Because of limited statistics, no cross sections are given 
          for the production of \antilambda\ particles on the W target.
          }  
\begin{center}
\setlength{\extrarowheight}{4pt}
\begin{tabular}{|c|c|c|c|c|} \hline
$\Delta p_t^2$ & C & Al &  Ti & W  \\ 
\hline 
     \multicolumn{5}{|c|}{$K^0_S$} \\ \hline
$0.0 - 0.2$ & $230.\pm 20._{-18.}^{+37}$ & $454.\pm 54._{-41.}^{+77.}$ 
          & $619.\pm53._{-50.}^{+99.}$ & $1901.\pm254._{-171.}^{+323.}$  \\
$0.2 - 0.4$ & $\ 94.\pm 10._{-\ 8.}^{+15.}$ & $191.\pm 28._{-17.}^{+32.}$ 
          & $275.\pm29._{-22.}^{+44.}$ & $\ 899.\pm141._{-\ 81.}^{+153.}$  \\
$0.4 - 0.6$ & $\ 49.\pm\ 7._{-\ 4.}^{+\ 8.}$ & $101.\pm 19._{-\ 9.}^{+17.}$ 
          & $156.\pm20._{-12.}^{+25.}$ & $\ 560.\pm\ 92._{-\ 50.}^{+\ 95.}$  \\
$0.6 - 0.8$ & $\ 28.\pm\ 5._{-\ 2.}^{+\ 4.}$ & $\ 64.\pm 14._{-\ 5.}^{+11.}$ 
          & $\ 99.\pm15._{-\ 8.}^{+16.}$ & $\ 229.\pm\ 57._{-\ 21.}^{+\ 39.}$  \\
$0.8 - 1.0$ & $\ 18.\pm\ 4._{-\ 1.}^{+\ 3.}$ & $\ 33.\pm 10._{-\ 3.}^{+\ 6.}$ 
          & $\ 51.\pm11._{-\ 4.}^{+\ 8.}$ & $\ 146.\pm\ 47._{-\ 13.}^{+\ 25.}$  \\
$1.0 - 1.2$ & $\ 14.\pm\ 3._{-\ 1.}^{+\ 2.}$ & $\ 29.\pm 10._{-\ 3.}^{+\ 5.}$ 
          & $\ 42.\pm10._{-\ 3.}^{+\ 7.}$ & $\ \ 75.\pm\ 34._{-\ \ 7.}^{+\ 13.}$  \\
\hline
     \multicolumn{5}{|c|}{\lambdazero} \\ \hline
$0.0 - 0.2$ & $58.\pm11._{-\ 8.}^{+10.}$ & $103.\pm23._{-14.}^{+20.}$ 
          & $\ 91.\pm34._{-12.}^{+15.}$ & $677.\pm200._{-\ 95.}^{+129.}$  \\
$0.2 - 0.4$ & $53.\pm\ 6._{-\ 7.}^{+\ 9.}$ & $\ 78.\pm11._{-11.}^{+15.}$ 
          & $167.\pm20._{-22.}^{+28.}$ & $327.\pm\ 81._{-\ 46.}^{+\ 62.}$  \\
$0.4 - 0.6$ & $32.\pm\ 4._{-\ 4.}^{+\ 5.}$ & $\ 54.\pm\ 8._{-\ 8.}^{+10.}$ 
          & $\ 94.\pm13._{-12.}^{+16.}$ & $364.\pm\ 60._{-\ 51.}^{+\ 69.}$  \\
$0.6 - 0.8$ & $17.\pm\ 3._{-\ 2.}^{+\ 3.}$ & $\ 33.\pm\ 5._{-\ 5.}^{+\ 6.}$ 
          & $\ 70.\pm10._{-\ 9.}^{+12.}$ & $249.\pm\ 45._{-\ 35.}^{+\ 47.}$  \\
$0.8 - 1.0$ & $15.\pm\ 3._{-\ 2.}^{+\ 3.}$ & $\ 27.\pm\ 5._{-\ 4.}^{+\ 5.}$ 
          & $\ 57.\pm10._{-\ 7.}^{+10.}$ & $263.\pm\ 45._{-\ 37.}^{+\ 50.}$  \\
$1.0 - 1.2$ & $\ 8.\pm\ 2._{-\ 1.}^{+\ 1}$ & $\ 14.\pm\ 3._{-\ 2.}^{+\ 3.}$ 
          & $\ 44.\pm\ 8._{-\ 6.}^{+\ 7.}$ & $120.\pm\ 31._{-\ 17.}^{+\ 23.}$  \\
\hline
     \multicolumn{5}{|c|}{\antilambda} \\ \hline
$0.0 - 0.2$ & $19.\ \pm 5.\ _{-1.\ }^{+4.\ }$ & $49.\pm11._{-\ 4.}^{+11.\ }$ 
          & $ 71.\pm17._{-\ 6.}^{+16}$ & $-$  \\
$0.2 - 0.4$ & $16.\ \pm 2.\ _{-1.\ }^{+3.\ }$ & $29.\pm\ 4._{-\ 2.}^{+\ 6.\ }$ 
          & $ 29.\pm\ 6._{-\ 2.}^{+\ 6.}$ & $-$  \\
$0.4 - 0.6$ & $\ 8.\ \pm 1.\ _{-0.5}^{+1.6}$ & $ 16.\pm\ 3._{-\ 1.}^{+\ 4.\ }$ 
          & $ 28.\pm\ 4._{-\ 2.}^{+\ 6.}$ & $-$  \\
$0.6 - 0.8$ & $\ 5.\ \pm 1.\ _{-0.3}^{+1.\ }$ & $ 16.\pm\ 2._{-\ 1.}^{+\ 4.\ }$ 
          & $ 20.\pm\ 4._{-\ 2.}^{+\ 4.}$ & $-$  \\
$0.8 - 1.0$ & $\ 4.6\pm 0.9_{-0.3}^{+1.\ }$ & $10.\pm\ 2._{-\ 0.8}^{+\ 2.\ }$ 
          & $10.\pm\ 3._{-\ 1.}^{+\ 2.}$ & $-$  \\
$1.0 - 1.2$ & $\ 2.4\pm 0.7_{-0.1}^{+0.5}$ & $\ 5.\pm\ 1._{-\ 0.4}^{+\ 1.\ }$ 
          & $\ 9.\pm\ 2._{-\ 1.}^{+\ 2.}$ & $-$  \\
\hline
\end{tabular}
\end{center} 
\label{tab:diffptall}
\end{table*}
%
\begin{table}[htb]
\caption{ The values of the parameter $B$ obtained by fitting the 
        differential cross section to the form
           ${d \sigma}/{dp^2_t} \propto exp(-B\cdot p^2_t)$,
           together with fit errors.
          Because of limited statistics, no value is obtained for 
          \antilambda\ particles produced on the W target.
         }
\begin{center}
\begin{tabular}{|l|c|c|c|c|} \hline
            &\multicolumn{4}{c|}{$B$\ \pumm} \\ \cline{2-5}
            & C &Al & Ti & W \\ \hline
\kzeros     & $3.3\pm0.2$ & $3.2\pm0.3$ & $3.0\pm0.2$ & $3.3\pm0.3$ \\
\lambdazero & $2.3\pm0.3$ & $2.0\pm0.3$ & $1.7\pm0.3$ & $1.3\pm0.5$ \\
\antilambda & $2.2\pm0.3$ & $2.0\pm0.3$ & $1.8\pm0.3$ & $    -    $ \\ \hline
\end{tabular}
\end{center}
\label{tab:bparam}
\end{table}
\clearpage

\end{document}